\documentclass[submitting]{nst}

\usepackage{subfigure,dcolumn}
\usepackage{epstopdf}
\usepackage[version=4]{mhchem}
\usepackage{upgreek}

\def\krpb {$^{86}$Kr +$^{208}$Pb}

\def\esym{$E_{sym}(\rho)$}

\def\pacab{${\rm PPAC 1\times 2}$}
\def\pacac{${\rm PPAC 1\times 3}$}

\def\esym {$E_{\rm sym}(\rho)$}

\begin{document}

\title{Reconstruction of Fission Events in Heavy Ion Reactions with CSHINE}

\author{Xinyue Diao}
\email[E-mail: ]{dxy17@mails.tsinghua.edu.cn}
\affiliation{Department of Physics, Tsinghua University, Beijing 100084,  China}

\author{Fenhai Guan}
\affiliation{Department of Physics, Tsinghua University, Beijing 100084,  China}

\author{Yijie Wang}
\affiliation{Department of Physics, Tsinghua University, Beijing 100084,  China}

\author{Yuhao Qin}
\affiliation{Department of Physics, Tsinghua University, Beijing 100084,  China}

\author{Zhi Qin}
\affiliation{Department of Physics, Tsinghua University, Beijing 100084,  China}

\author{Dong Guo}
\affiliation{Department of Physics, Tsinghua University, Beijing 100084,  China}

\author{Qianghua Wu}
\affiliation{Department of Physics, Tsinghua University, Beijing 100084,  China}

\author{Dawei Si}
\affiliation{Department of Physics, Tsinghua University, Beijing 100084,  China}

\author{Xuan Zhao}
\affiliation{Department of Physics, Tsinghua University, Beijing 100084,  China}

\author{Sheng Xiao}
\affiliation{Department of Physics, Tsinghua University, Beijing 100084,  China}

\author{Yaopeng Zhang}
\affiliation{Department of Physics, Tsinghua University, Beijing 100084,  China}

\author{Xianglun Wei}
\affiliation{Institute of Modern Physics, Chinese Academy of Sciences, Lanzhou 730000,  China}

\author{Haichuan Zou}
\affiliation{School of Nuclear Science and Technology, Lanzhou University, Lanzhou 730000,  China}

\author{Herun Yang}
\affiliation{Institute of Modern Physics, Chinese Academy of Sciences, Lanzhou 730000,  China}

\author{Peng Ma}
\affiliation{Institute of Modern Physics, Chinese Academy of Sciences, Lanzhou 730000,  China}

\author{Rongjiang Hu}
\affiliation{Institute of Modern Physics, Chinese Academy of Sciences, Lanzhou 730000,  China}

\author{Limin Duan}
\affiliation{Institute of Modern Physics, Chinese Academy of Sciences, Lanzhou 730000,  China}

\author{Artur Dobrowolski}
\affiliation{Uniwersytet Marii Curie Sk{\l}odowskiej, Katedra Fizyki Teoretycznej, Lublin 20031, Poland}

\author{Krzysztof Pomorski}
\affiliation{Uniwersytet Marii Curie Sk{\l}odowskiej, Katedra Fizyki Teoretycznej, Lublin 20031, Poland}

\author{Zhigang Xiao}
\affiliation{Department of Physics, Tsinghua University, Beijing 100084,  China}

\begin{abstract}
 
 We report the reconstruction method of the fast fission events in 25 MeV/u $^{86}$Kr +$^{208}$Pb reactions at the  Compact Spectrometer for Heavy IoN Experiment (CSHINE). The fission fragments are measured by three  large-area parallel plate avalanche counters, which can deliver the position and the arrival timing information of the fragments. The start timing information is given by the radio frequency of the cyclotron. Using the velocities of the two fission fragments, the fission events are reconstructed. The  broadening of both the velocity distribution and the azimuthal difference of the fission fragments decrease with the folding angle, in accordance with the picture that fast fission occurs. The  anisotropic angular distribution of the fission axis also reveals consistently the dynamic feature the fission events.
  
\end{abstract}


\maketitle

\section{Introduction}\label{sec.I}

One of the purposes of studying the heavy ion reactions (HIR) is to infer the properties of nuclear equation of state (EOS), which is an essential input in modeling and computing the evolution and properties of neutron stars and their merging \cite{Liba2021,Ligo2017,Ligo2018}.  The isovector sector of the nuclear EOS, namely the density behavior of the symmetry energy \esym, has been a long-standing  open question in nuclear physics. At Fermi energies,  some observables have been identified to constrain \esym~ near normal density, including isospin diffusion,  dipole polarizability and particle emissions  etc \cite{Tsang2004,Chenlw2005,Tsang2001,Chenlw2010,Zz2014,Tsang2009,Zy2017}.
 Very recently, the PREX II experiment has reported the result of the neutron skin thickness, which yields a stiff \esym~ in tention with previously existing  constraints \cite{Prex2,Reed2021}.  At supra-saturation densities, many experiments  have been progressed greatly to measure the charged pion ratios or collective flow in heavy ion collisions to probe \esym~ \cite{Sprit2021,Yongjia2020}. An External-target Experiment on  HIRFL-CSR (CEE) is under construction for the studies on this direction  \cite{LLM2017,LLM2020}.

Nuclear fission is a large-amplitude collective motion mode involving up to hundreds of nucleons. Recently the studies on nuclear fission have been revived for its significance in both nuclear physics and astrophysics. In the stellar environment, the abundance of the nuclide in $A\approx 160$ region is significantly influenced by the recycling of the fission products \cite{Lorusso2015,Nishimura2012,Suzuki2012}. Theoretically, statistical fission has been described well by microscopical theories and various phenomenological approaches \cite{PBR96,SCH99,JB2007,LB2011,Zhanghf2014,Tanimura2017,TZL17,WY2018,WY2018-2,Pomorski2021, Pavel2021,Zhengh2018,Guol2018}. When the excitation energy or the angular momentum becomes much high, as achieved in HIRs well above the Coulomb barrier, the fission barrier tends to vanish. As a consequence,  the fission time scale becomes shorter by a factor of 10 to 100 and the variance of the mass asymmetry increases significantly (the mass asymmetry $\eta$ of the two fragments can be larger than 0.6), compared to the statistical fission \cite{Greg82, Greg82t, Gla83, Leray84, Zheng84}. In this case, the dynamic feature of the fission process is of significance and the transport models have been successfully applied to describe the fast fission process \cite{Wen13, Russ11,Riz11,TL09,TO11,LTQ13,TW08,LTO13,WT11}.Time-dependent Hartree-Fock theory can also give an excellent description of fast fission starting with large deformation \cite{God2015}.  Experimentally, fast fission, usually termed as dynamic fission because of its vanishing fission barrier and short time scale, has  been investigated in various systems in the last three decades \cite{Bocage2000,Filippo2005,Filippo2012,Pagano2018,Piantelli2020,WRS2014,Casini1993}.

The topic of fast fission with the simultaneous emission of the particles deserves further investigations because the fissioning system provides an appropriate laboratory to probe \esym. The connection between the studies of fast fission and \esym~ has been recently  established  by the simulations with improved quantum molecular dynamics (ImQMD) \cite{Wuqh2019,Wuqh2020}. It is suggested that fast fission process following HIR carries the effect of \esym~ and provides sensitive probes, because of the formation of the low-density and neutron-rich neck and of larger surface of two fragments compared to the non-fission process  \cite{Wuqh2020,Pei2020}.  The isospin content of the light particles emitted from the fast fission events has been used to probe \esym~ experimentally \cite{Zy2017}.  

In order to conduct the experimental studies of the fast fission and the coincident emission of light charged particles (LCP) and intermediate mass fragments (IMFs), a compact spectrometer for heavy ion experiment (CSHINE) has been built \cite{Guanfh2021,Wangyj2021}.  While the LCPs and IMFs are measured using the silicon strip detector telescopes (SSDT) \cite{Guanfh2022,Wangyj2022}, the fission fragments are measured by the parallel plate avalanche counters (PPACs) \cite{Weixl2020}. 

In this paper, we present the measurement of fast fission in \krpb~ reactions with CSHINE in its second phase.   After a brief introduction to the phase-II setup of CSHINE  in section II, the reconstruction of velocity of the FF will be introduced in Section III and the dynamic feature of fast fission are presented in Section IV. Section V is the summary.

\section{CSHINE detector system and the experimental setup}\label{sec.II}

\begin{figure}[!htb]
\includegraphics[width=.45\textwidth]{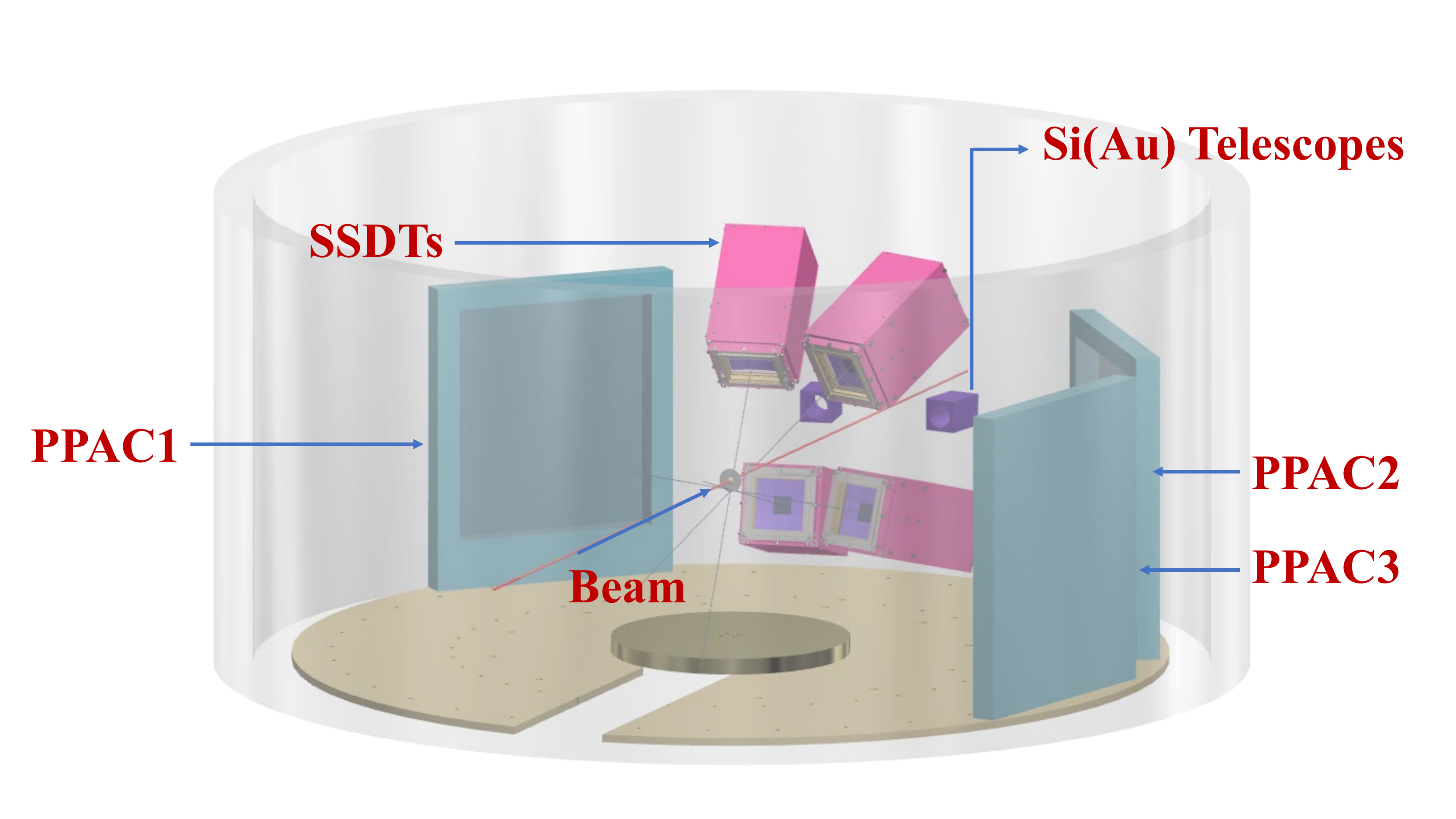}
\caption{The schematic view of CSHINE Phase-II.}
\label{cshine}
\end{figure}

The beam experiment was performed at the Radioactive Ion Beam Line I  (RIBLL1) on the Heavy Ion Research Facility in Lanzhou (HIRFL), China. The $^{208}$Pb target of 1 mg/cm$^2$ areal density was bombarded with 25 MeV/u $^{86}$Kr beam.  
The charged  reaction products are measured using CSHINE, which are installed in a large scattering chamber located at the final focal plane of RIBLL1.  In the current experiment, three PPACs are installed for fission fragment measurements to reconstruct the reaction geometry. Besides, four SSDTs  are installed covering the polar angle range $10^\circ<\theta_{\rm lab}<60^\circ$.  The SSDT is a three-layer detector with a single-sided  SSD (SSSSD) for $\Delta E_1$ as layer 1, a  double-sided SSD (DSSSD) for $\Delta E_2$ as layer 2 and a  $3\times3$ CsI (Tl) array for residual energy measurements as layer 3.  Both of the SSSSD and the DSSSD are the BB7 type (2 mm strip width, 32 strips each side) from MICRON Company.  Each CsI (Tl) crystal is a square pyramid with a dimension of  $23\times23$ mm$^2$ of the front side, $27\times27$ mm$^2$ of the rear side and 50 mm of the height. The  Photo Diode (HAMAMATSU  S3204) is used to read out the signal from CsI.  Fig. \ref{cshine} presents the detector setup of CSHINE in the experiment. For the details and the performance  of CSHINE, one can refer to \cite{Wangyj2021,Guanfh2021}. Table 1 presents the distance $d$ from the center of each detector to the target, the polar angle $\theta$, the azimuthal  angle $\phi$ and the sensitive area $S$ of the SSDTs and the PPACs in the experiment. The thicknesses of the $\Delta E_1$  and $\Delta E_2$  for each SSDT are also listed.

\begin{table} [!htb]
	\label{tab:cee_para}
	\caption{Geometric parameters of PPACs and SSDTs.}
	\begin{center}
		\begin{tabular}
			{ccccccc}\toprule[0.65pt]  
			\hline 
			Detector & $d$(mm) & $\theta(^{\circ})$ & $\phi(^{\circ})$ & $S$(mm$^2$)  &$\Delta E_1$ ($\mu$m) & $\Delta E_2$ ($\mu$m) \\
			\hline
			SSDT1  & 315.5 & 18 & 302 & 64$\times$64  & 304 & 1010\\
			SSDT2  & 275.5 & 25 & 218 & 64$\times$64  & 305 & 1008\\
			SSDT3  & 275.5 & 31 & 126 & 64$\times$64  & 110 & 526\\
			SSDT4  & 215.5 & 51 & 81 & 64$\times$64  & 70 & 306\\
			PPAC1  & 427.5 & 50 & 0 & 240$\times$280 &$-$&$-$\\
			PPAC2  & 427.5 & 55 & 180 & 240$\times$280&$-$ &$-$ \\
			PPAC3  & 427.5 & 100 & 180 & 240$\times$280& $-$&$-$\\

			\hline \bottomrule[0.5pt]
		\end{tabular}
	\end{center}
\end{table}
\vspace*{-2mm}

\begin{figure}[!htb]
\includegraphics[width=.45\textwidth]{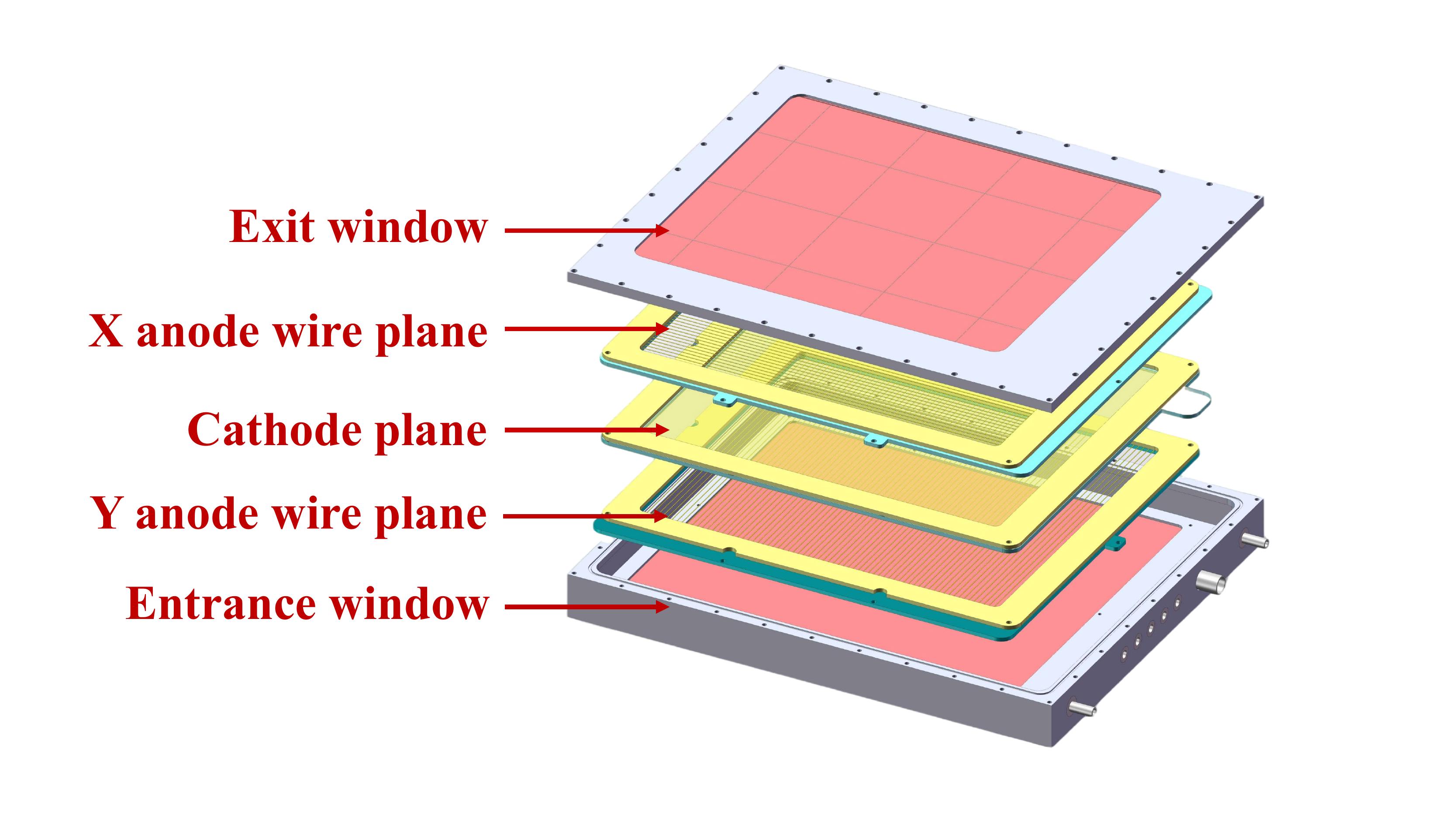}
\caption{(Color online) The schematic view of  PPAC as the fission fragment detector.}
\label{ppac}
\end{figure}

The fission fragment detector,  PPAC, is a kind of   multi-wire chamber working in the region of limited proportionality. The signals induced by the incident fragments on an individual wire of the anode plane, either X or Y, are transferred through a delay line to both ends. The time delay of the two signals $X_1$ and $X_2$ ($Y_1$ and $Y_2$)  with respect to the signal collected on the cathode plane, which delivers the timing information,  gives the X (Y) position of the hit in the sensitive area. Fig. \ref{ppac} presents the schematic view of the mechanics of the PPAC.  The total thickness of the sensitive gas volume is about 2 cm. The PPACs were operated with 4.5 mbar isobutane at 465 V voltage. Under this condition, the fission fragments can be recorded with efficiency above 95\% but the LCPs and IMFs are suppressed.

\begin{figure}[!htb]
\includegraphics[width=.45\textwidth]{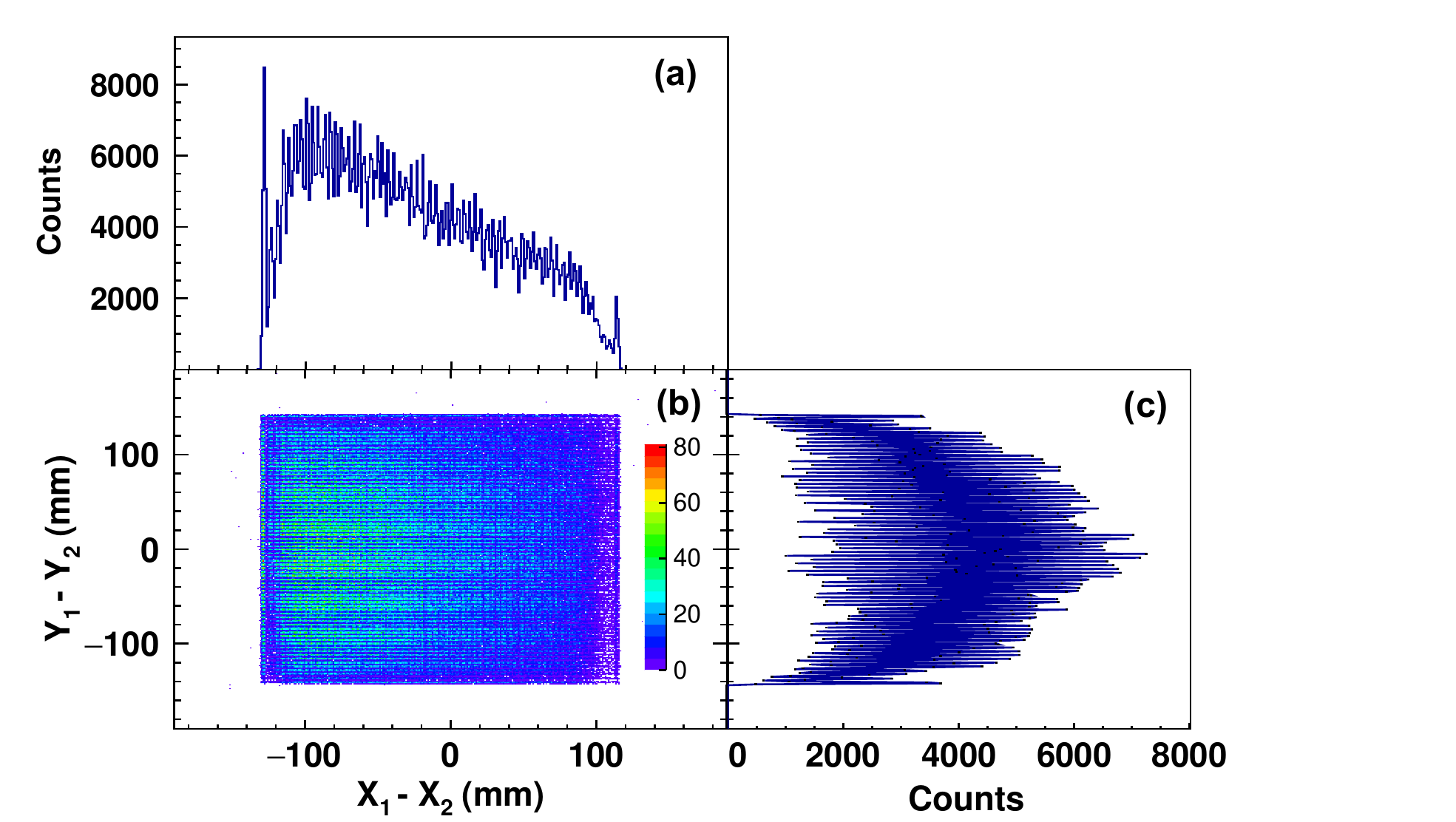}
\caption{(Color online) The scattering plot of  $Y_1-Y_2$ vs. $X_1-X_2$ of PPAC1.  Panels (a) and (c) are the projections of the two dimensional plot of X-Y in panel (b).}
\label{ppac_xy}
\end{figure}

Fig. \ref{ppac_xy} (b) shows a two-dimensional histogram of $Y_1-Y_2$ $vs$ $X_1-X_2$  for PPAC1 as an example. The projections to X and Y direction are plotted in Fig. \ref{ppac_xy} (a) and (c), respectively. A good performance in timing, corresponding to good position resolution, manifests itself in the sharp boundary for the two-dimensional distribution and the well-separated individual peaks on the projections. The distance of the neighboring wires is 4 mm, and there are 61 peaks and 71 peaks in Fig. \ref{ppac_xy} (a) and (c), respectively. The time resolution of $\sigma_{\rm T} = 300$ ps and the position resolution of $\sigma_{\rm r}=1.35$ mm can be derived from the data. The overall performance of the PPACs can be found in  \cite{Weixl2020}.

SSDTs are used to measure the LCPs and IMFs in coincidence with the fission fragments.  To reduce the total number of electronics, each neighboring strips are merged in one channel, correspondingly the granularity is reduced.  Multi tracks may fire each SSDT. In order to reconstruct the tracks in the SSDTs, a novel algorithm has been developed with special care on the charging sharing effect.  More than 80\% of the hits in all layers of the SSDTs can be recognized and assigned to certain tracks.  For details, one can refer to \cite{Guanfh2022}.  

The trigger system of CSHINE was designed for both beam experiment and calibration.  The timing signals of PPACs were discriminated by CF8000 and logically calculated by a CO4020 to generate the PPAC inclusive signals and PPAC two body coincidence signals. The logic hit signals of SSDTs were extracted by the front-side of the DSSSD ($\Delta E_2$) with the discrimination of MSCF-16, which generates an analog Multi-Trig signal proportional to the number of fired strips in the same module (16 channels). Both inclusive and exclusive logic signals can be generated by discriminating the Multi-Trig signal at different threshold settings.  In beam experiment, the trigger signal contains SSD two body events, PPAC two body events  and  the coincidence of PPAC two body with SSD one body events. Besides, the inclusive trigger for every individual detector were also constructed, optionally turned on for the detector calibration before or after the beam data-taking. For details, one can refer to  \cite{Guanfh2022}.

\section {Reconstruction of the Velocity} \label{sec.III}

We concentrate on the reconstruction of the fission events. The flying path of the fission fragments (FFs) can be well determined by the PPACs delivering good position information. The velocity of each FF, on the other hand, is derived from the timing information. In our experiment, the starting time information are provided by the radio frequency (RF) of the accelerator.   The RF signal, usually a signal in  sinusoidal form, is discriminated by CF8000 module and input to the time-digital-converter (TDC). Generally, for a particle firing a given detector, the time of flight (\textit{TOF}) is written as

\begin{equation}
	TOF = t_{\rm det} - t_{\rm RF} - C_{\rm det}   
\end{equation}

where  $t_{\rm det}$ and $t_{\rm RF}$ are the time signals of the detector and the RF, recorded by the corresponding TDC channels, respectively.  The unit of both signals are calibrated in ns using precise time calibrator.   $C_{\rm det}$  is a constant representing a fixed delay  in electronics. The velocity is then computed by 

\begin{equation}
	v =\frac{L}{TOF}   
\end{equation}

where $L$ is the length of the flight path from the target to the hit position where the particle fires on the detector.

\begin{figure}[!htb]
\includegraphics[width=.45\textwidth]{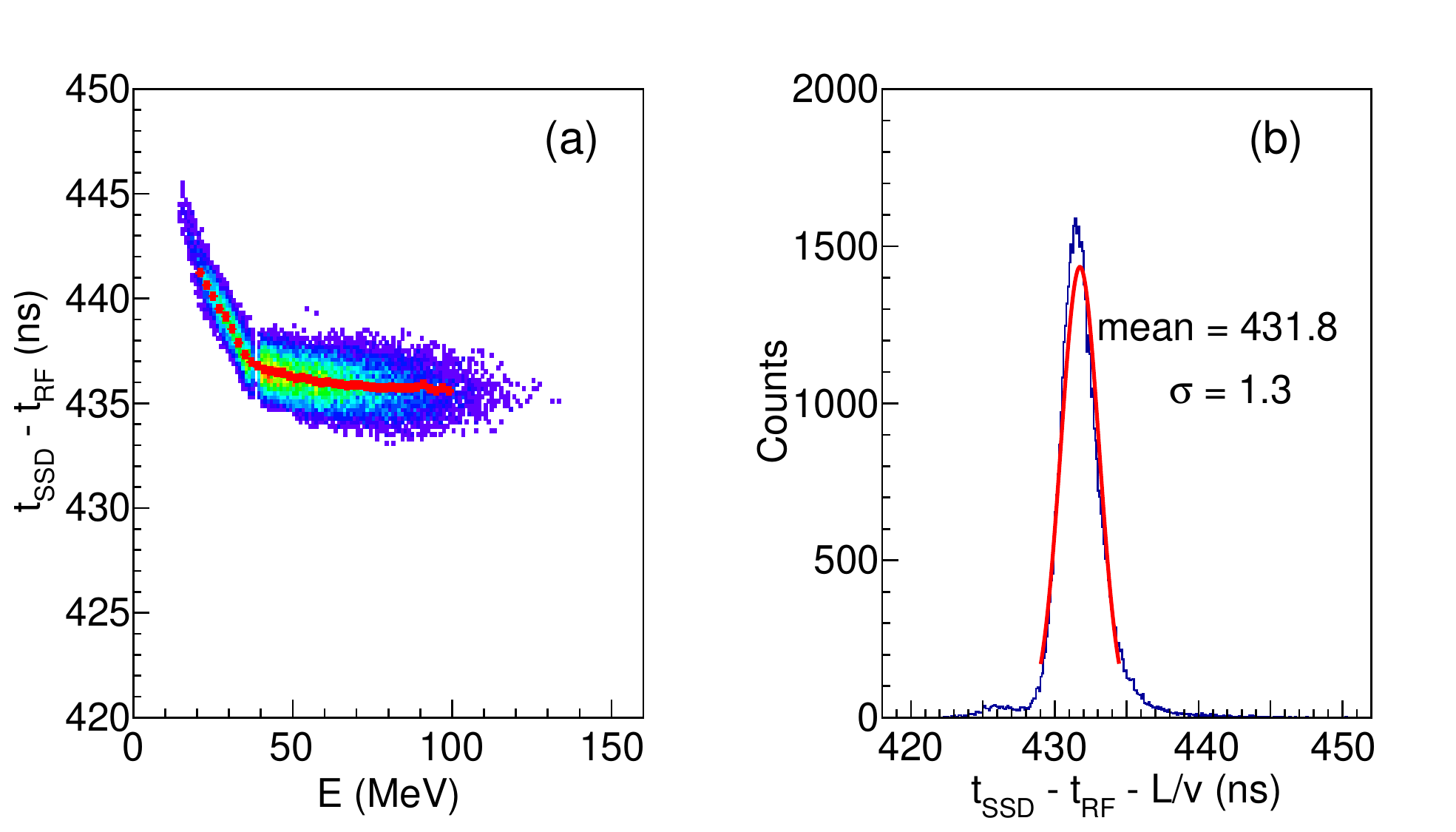}
\caption{(a) Correlation of the \textit{TOF} signal and the energy of $\alpha$ particles recorded by SSDT3, and (b) The distribution of the time difference between the TDC measurement  and the calculated value from the calibrated energy.}
\label{rf_check}
\end{figure}

 In order to verify the validity of the above method to measure the \textit{TOF}, we use the calibrated  $\alpha$ particles, where the velocity is alternatively derived from the energy measured in SSDT3.  Fig. \ref{rf_check} (a) presents the correlation between the  $\alpha$ energy and the \textit{TOF} derived by eq. (1). The theoretical curve fitting the  \textit{E-TOF} profile applies a constant $C_{\rm det}=431.8$ ns. Panel (b) presents the difference between the \textit{TOF} measured by TDC and the value $L/v(E_\alpha)$, with $v$ being derived from the energy $E_\alpha$ and $L$ is the distance from the target to  hit position in SSDT3. The width of 1.3 ns  is obtained by the Gaussian fitting. Subtracting the contribution of the energy uncertainty, the resolution  of about 1.0 ns of the \textit{TOF} is  obtained for SSDT. For the  FF measured in PPACs, the \textit{TOF} resolution will be comparable because the timing resolution of PPACs is 300 ps, slightly better than the SSDT.   

\begin{figure}[!htb]
\includegraphics[height=8cm,width=.35\textwidth]{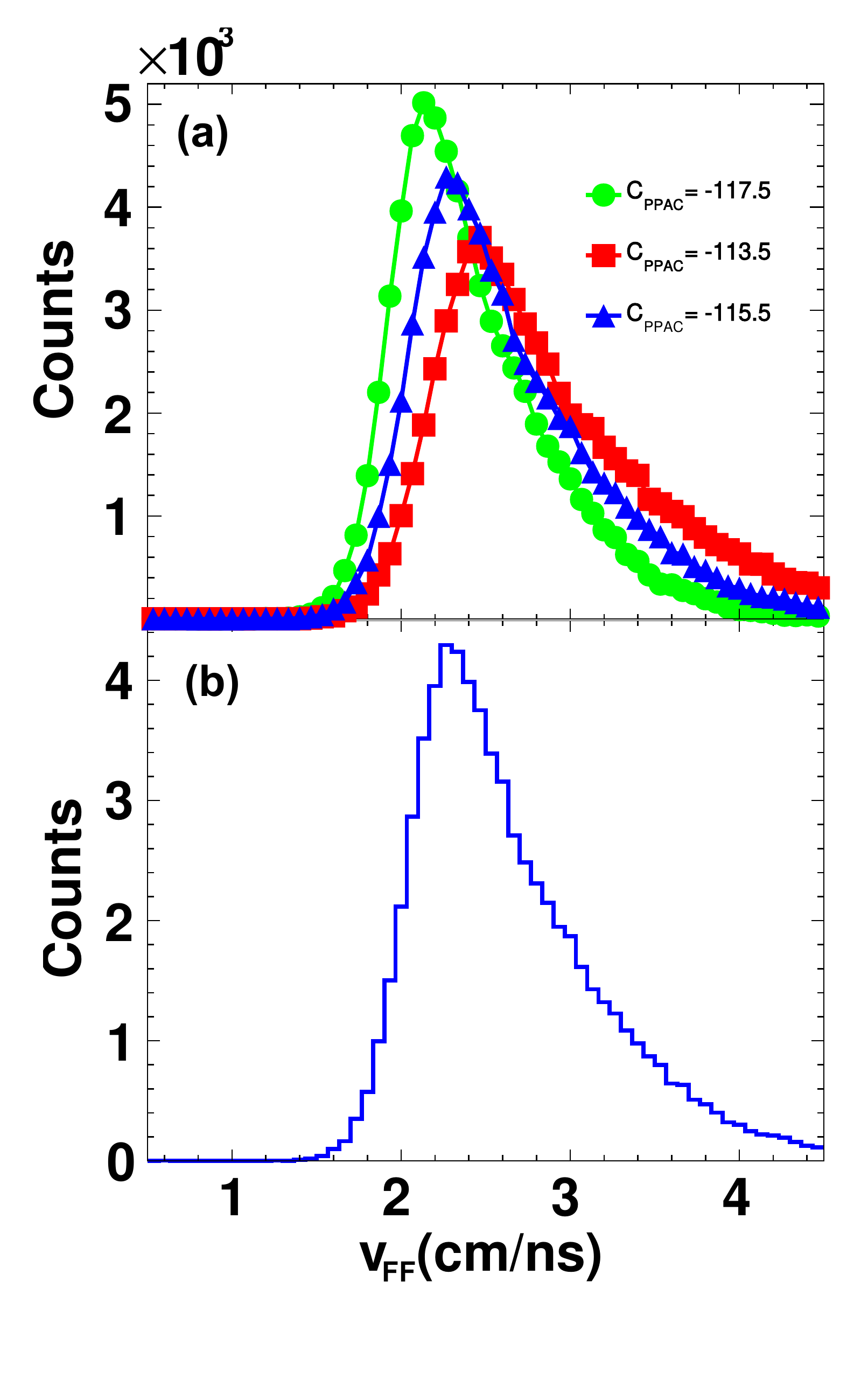}
\caption{The relative velocity distribution $v_{\rm FF}$  of the two fission fragments with different parameter setting of $C_{\rm PPAC}$.}
\label{t_const}
\end{figure}

With the validity eq. (1), we can discuss the  \textit{TOF} of the FFs.  Since PPACs can not identify the charge or mass of the FFs, nor the  total kinetic energy, one relies on the determination of the velocities which requires the \textit{TOF} information and the lengths of the flight path $L$.  In eq. (1),  the relative difference of the delay constants, $C_{\rm ppac1}-C_{\rm ppac2}$, between two PPACs can be adjusted to zero using pulser  prior to the experiment and the systematic uncertainty can be well controlled within  $\pm 2$ ns. However, the absolute value of $C_{\rm ppac}$ of each individual PPAC can not be determined like in the SSDT, because the particle type and the total energy is unknown. To overcome this difficulty, we use the Viola systematics that the relative velocity of the FFs is averagely $2.4 ~{\rm cm/ns}$ \cite{Viola} . And hence, by tuning the constants $C_{\rm ppac}$, one can  optimized the value at  $\left<v_{\rm FF}\right>=2.4 ~{\rm cm/ns}$.  Fig.  \ref{t_const} presents the distribution of  $v_{\rm FF}$  at different delay constants. One can readily see that varying the delay constant by 1 ns, the peak position of  $v_{\rm FF}$ moves significantly.  In our experiment, $C_{\rm ppac}=115.5$ ns is optimized, and the corresponding distribution of $v_{\rm FF}$  is plotted in panel (b) . Known from transport model calculations, the variation of $\left<v_{\rm FF}\right>$ is better than $0.1~{\rm cm/ns}$,    a systematic uncertainty of 2 ns of the \textit{TOF} of FFs is estimated, on which the conclusion of  the following analysis will not changed. We recall that  the coincident events recorded by  PPAC1 with PPAC2 (marked by ${\rm PPAC 1\times 2}$) or with PPAC3 (marked by ${\rm PPAC 1\times 3}$) are the fission fragments , since the HV condition  is set that the response of PPAC to energetic  LCPs and IMFs are totally suppressed, given that the energy loss of these particles is lower than FFs by more than one order of magnitude. In addition, the correlation between a heavy projectile-like fragment (PLF) and a target-like fragment (TLF) is out of the current geometrical coverage. 

\section{Results and Discussions}\label{sec.III}

\begin{figure}[!htb]
\includegraphics[width=.45\textwidth]{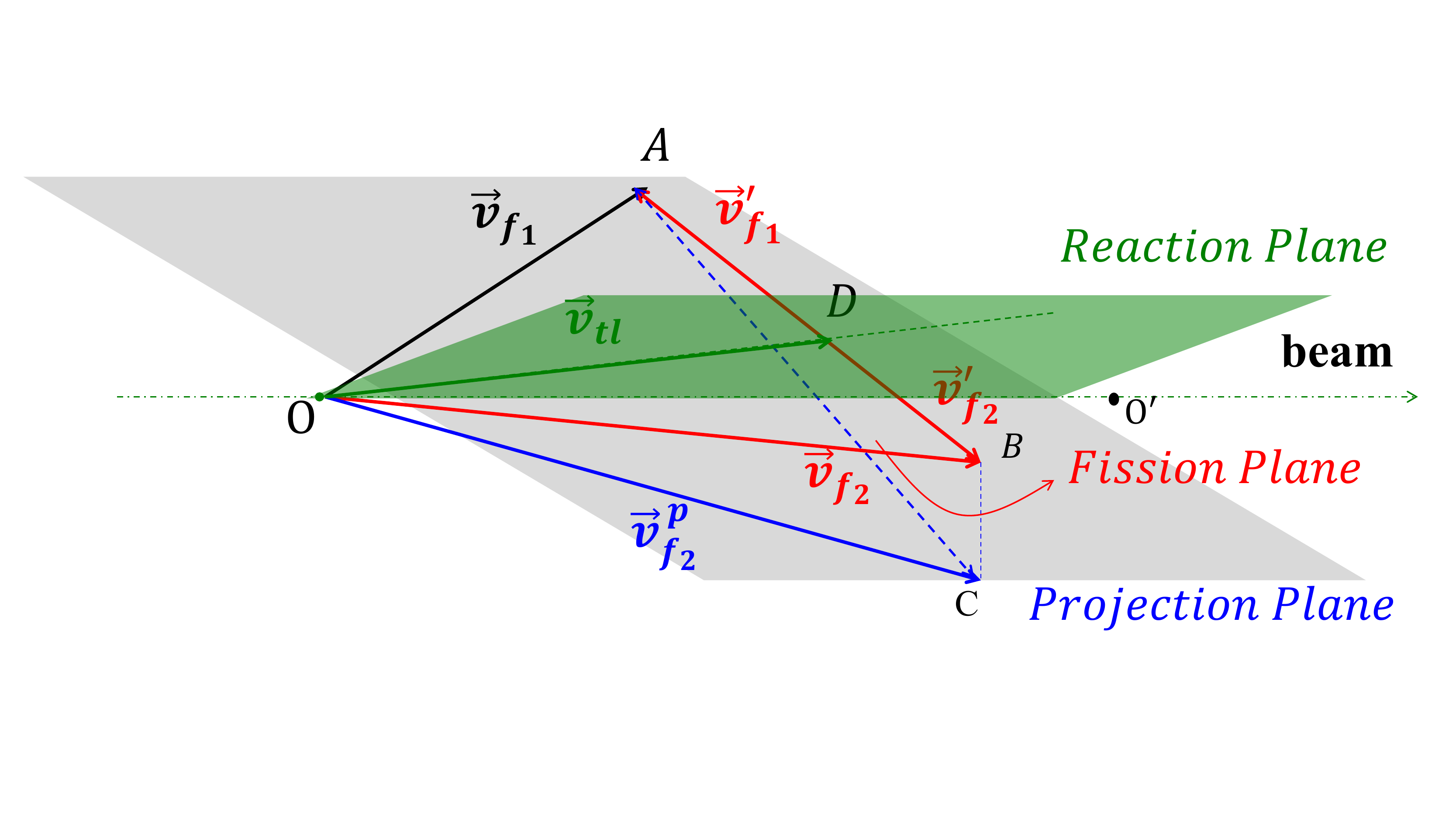}
\caption{The geometric diagram of the velocities of the two  fission fragments from TLF.}
\label{fission_vector}
\end{figure}

Before discussing  the reconstruction of the fission events, we first define the kinetics of the fission. In the picture of incomplete fusion, a heavy TLF is formed in the fusion of part of the projectile and the target. The fraction of the  momentum of the projectile transferred to the TLF is called linear momentum transfer (LMT). With a certain probability depending on the  total angular momentum of the reaction system, the TLF may undergo fission or fast fission in competition with the emission residue channel. For the fission events, Fig. \ref{fission_vector} presents the kinetic geometry of the TLF fission event. As shown, the origin point O is the target nucleus in the laboratory system, the vector OO$^{'}$ represents the direction of the beam. The velocity vectors $\vec{v}_{\rm f_1}$ and $\vec{v}_{\rm f_2}$ of the two fission fragments in the laboratory system are represented by OA and OB.   $\vec{v}_{\rm tl}$ is the velocity of the TLF, and the velocities of the two fragments in the center-of-mass system of the fissioning TLF are represented by $\vec{v}^{\,'}_{\rm f_1}$ and $\vec{v}^{\,'}_{\rm f_2}$ sitting back-to-back collinearly. Here, we define plane OAB as the fission plane, plane OO$^{'}$A as the projection plane, and the reaction plane is defined as plane DOO$^{'}$.

\begin{figure}[!htb]
\includegraphics[width=.45\textwidth]{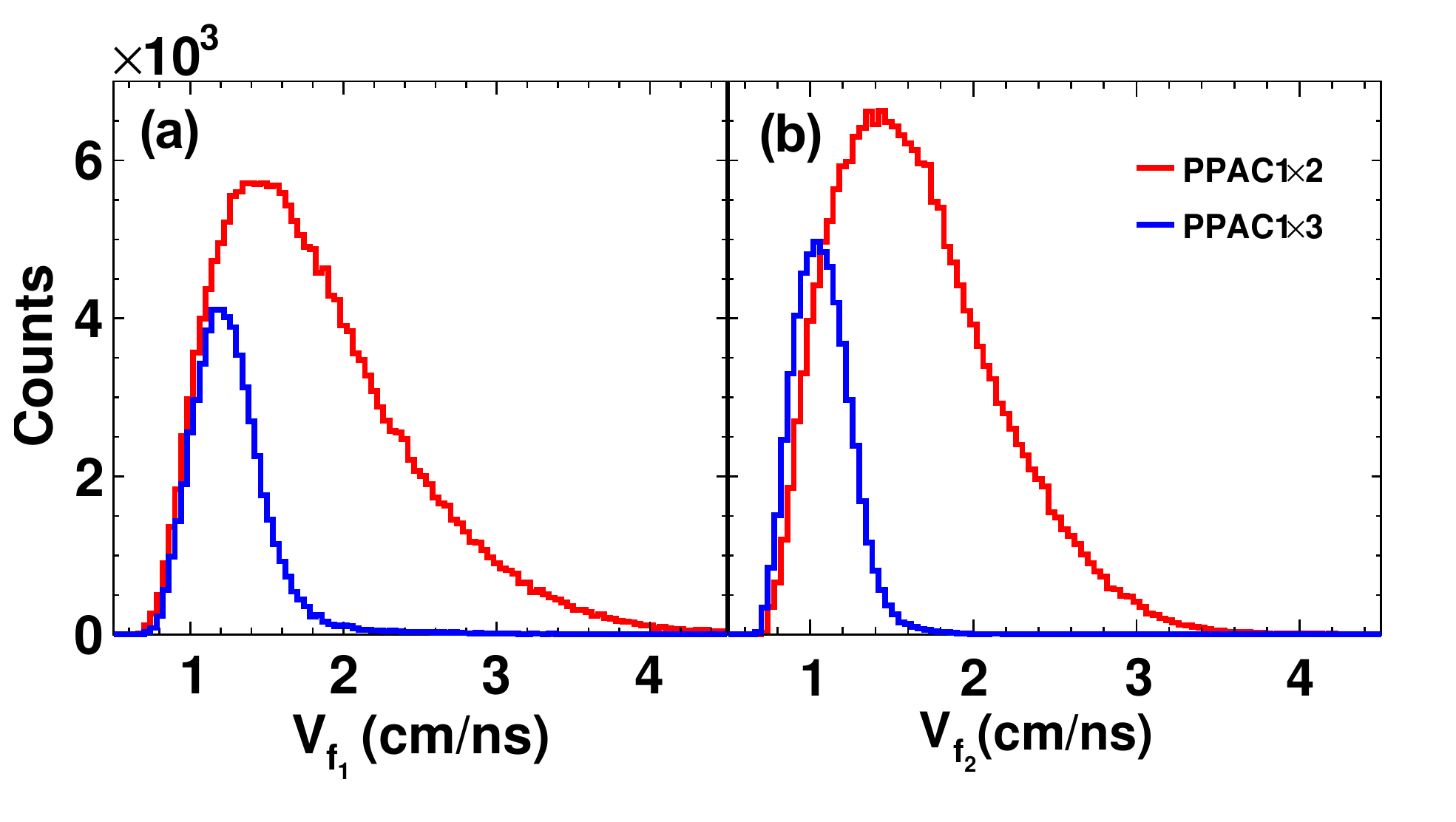}
\caption{The velocity distribution of the two fission fragments in \pacab~ and \pacac~ events. }
\label{v_rec}
\end{figure}

Once the  \textit{TOF} is determined, the velocity of the FFs can be computed event-by-event using the hit positions of the FFs in PPACs. Then the whole fission event can be reconstructed. Fig. \ref{v_rec} presents the distribution of the velocity of the fission fragments in \pacab~ and \pacac~ events, respectively.  Here the $v_{\rm f_1}$ is for the FF recorded in PPAC1 and   $v_{\rm f_2}$  is for the FF in PPAC2 (PPAC3) in \pacab~ (\pacac) events. It is shown that for the \pacab~ events, since the two PPACs are nearly symmetric w.r.t the beam, the distributions of  $v_{\rm f_1}$ and  $v_{\rm f_2}$ are very similar. Meanwhile,  high velocity tail is evidently presented  in the \pacab~ events and the velocity spectra are wider than that in \pacac~ events. This component is mainly due to the events with smaller folding angle corresponding to larger LMT, as will be discussed below. 

\begin{figure}[!htb]
\includegraphics[width=.45\textwidth]{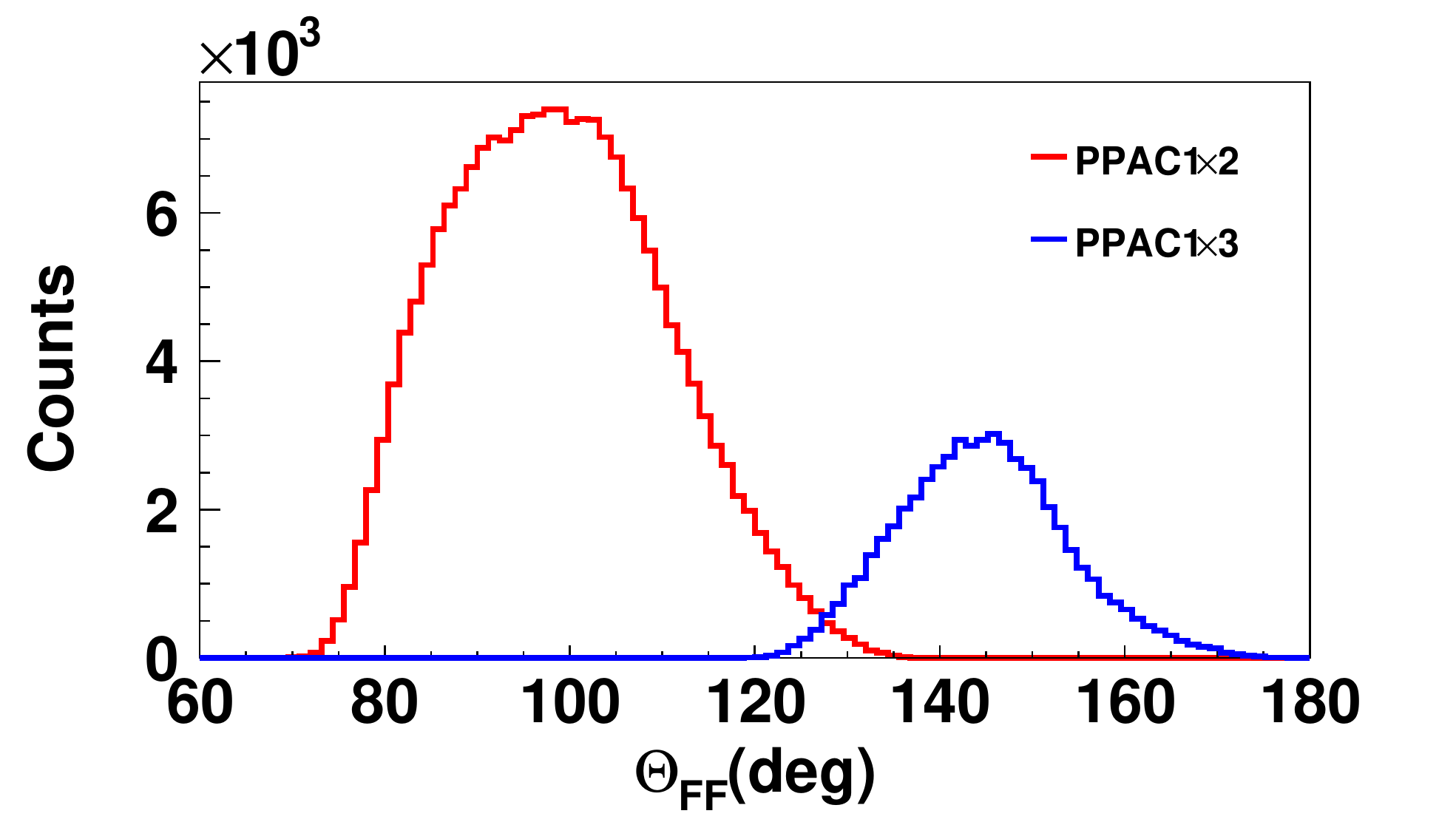}
\caption{(Color online) The folding angle distribution.}
\label{folding}
\end{figure}

After the velocities of the two FFs are determined, the folding angle method can be applied to calculate the LMT of the reaction. Recalling the definitions  in Fig. \ref{fission_vector}, the folding angle $\Theta_{\rm FF}$ is defined as the angle $\angle AOC$,  spanned by the projection of the velocity vectors of the  FFs on the projection plane. Clearly, given the velocities  $\vec{v}_{\rm f_1}$ and  $\vec{v}_{\rm f_2}$, the folding angle depends on the velocity of the TLF $v_{\rm tl}$, i.e., the larger $v_{\rm tl}$, the smaller  $\Theta_{\rm FF}$. Fig. \ref{folding} presents the distribution of the folding angle $\Theta_{\rm FF}$. The coincident events of \pacab~ distribute in the range between $70^{\circ}-120^{\circ}$ with the peak situating at about $95^{\circ}$, corresponding to a larger LMT (red), while the  \pacac~ events sit in the range of $120^{\circ}-170^{\circ}$  corresponds to a smaller LMT (blue).  The valley between the two components  is simply due to the deficiency caused by the  gap between PPAC2 and PPAC3, and the efficiency arising from the incomplete azimuthal coverage is not corrected in the plot.

\begin{figure}[!htb]
\includegraphics[height=10cm,width=.35\textwidth]{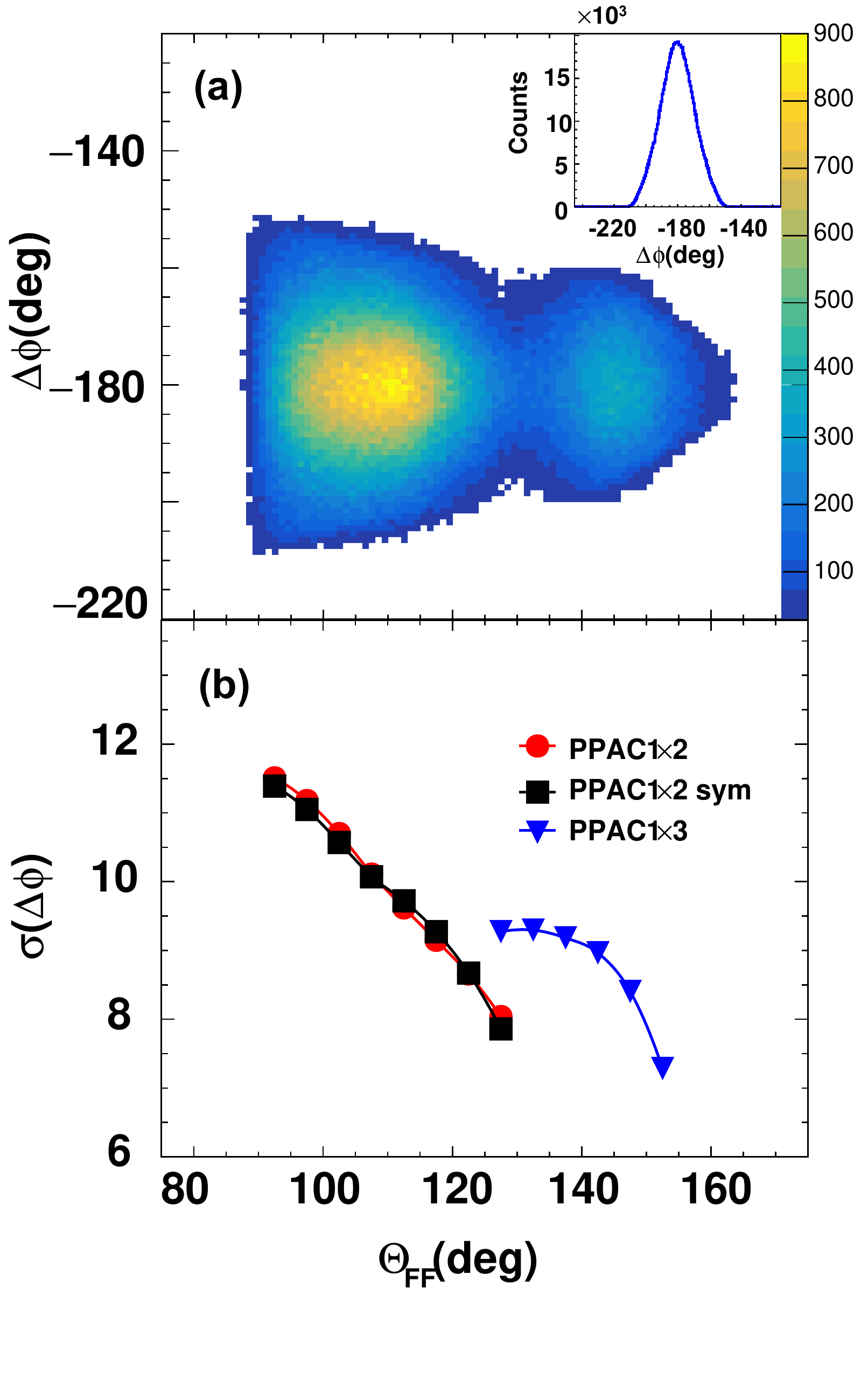}
\caption{(Color online) (a) The azimuthal angle difference $\Delta \phi$ as a function of the folding angle  $\Theta_{\rm FF}$ of the fission fragments. The inset displays the projection distribution of $\Delta \phi$. (b) $\sigma(\Delta \phi)$ as a function of $\Theta_{\rm FF}$.}
\label{d_phi}
\end{figure}

As the fission geometry is determined by the two velocity vectors, it is of interest to look at the planarity of the fission events.  Fig. \ref{d_phi} (a)  present the azimuth correlation of the two FFs by scattering plot between the azimuthal angle difference $\Delta \phi$ and the folding angle $\Theta_{\rm FF}$. It is clearly shown that for the FFs from both the central and peripheral reactions, the most probable value situates at  $\Delta \phi=180^\circ$ following the picture that the system undergoes a  binary decay. Here we note that the  $\Delta \phi$ is a directly measurable quantity relying on no assumption. The inset displays the projection distribution of $\Delta \phi$, a standard deviation of about $\sigma(\Delta \phi)\approx10^\circ$ is derived. Such broadening suggests that the emission of LCPs or IMFs may change the flight direction of the FF and smear the back-to-back feature of the fission event.  In order to see its evolution with the violence of the reaction,   we plot in panel (b) the standard deviation of azimuthal angle  $\sigma(\Delta \phi)$ as a function of folding angle $\Theta_{\rm FF}$.   It is shown that  $\sigma(\Delta \phi)$ decreases with  $\Theta_{\rm FF}$ in the whole range , with an exception near $\Theta_{\rm FF}\approx130^\circ$, where a discontinuity appears due to the gap between PPAC2 and PPAC3.  To exclude the possible reason that this trend originates from the asymmetry of the geometrical locations of the PPACs, we restrict further the analysis on the events with the two FFs flying symmetrically to the beam direction, i.e., with the condition of $\theta_1=\theta_2$, where $\theta_i$ is the polar angle of the $i^{\rm th}$ fragment. The result is depicted by the black squares, where the bin width of each $\Theta_{\rm FF}$ is $\pm 2.5^\circ$. This condition is applicable  only for the \pacab~ fission events because these two PPACs are placed in approximate left-right symmetry with respect to the beam line. Clearly seen, the data points with the symmetry condition are sitting on top of those without the condition, suggesting that the decreasing trend of $\sigma(\Delta \phi)$ as function of $\Theta_{\rm FF}$  is truly due to the reaction violence. Since the post-scission particle emission changes the velocity of the FF due to recoil effect, the trend suggests that in the fission following the intermediate energy heavy ion reactions, there is sufficient excitation energy left at the scission point depending on LMT. In the reactions with larger LMT, more excitation energy is left and released through the particle emission in the post-scission stage. It is consistent with the picture of fast fission, instead of statistic fission in which the excitation energy is nearly depleted at the scission point.

\begin{figure}[!htb]
\includegraphics[height=8cm,width=.35\textwidth]{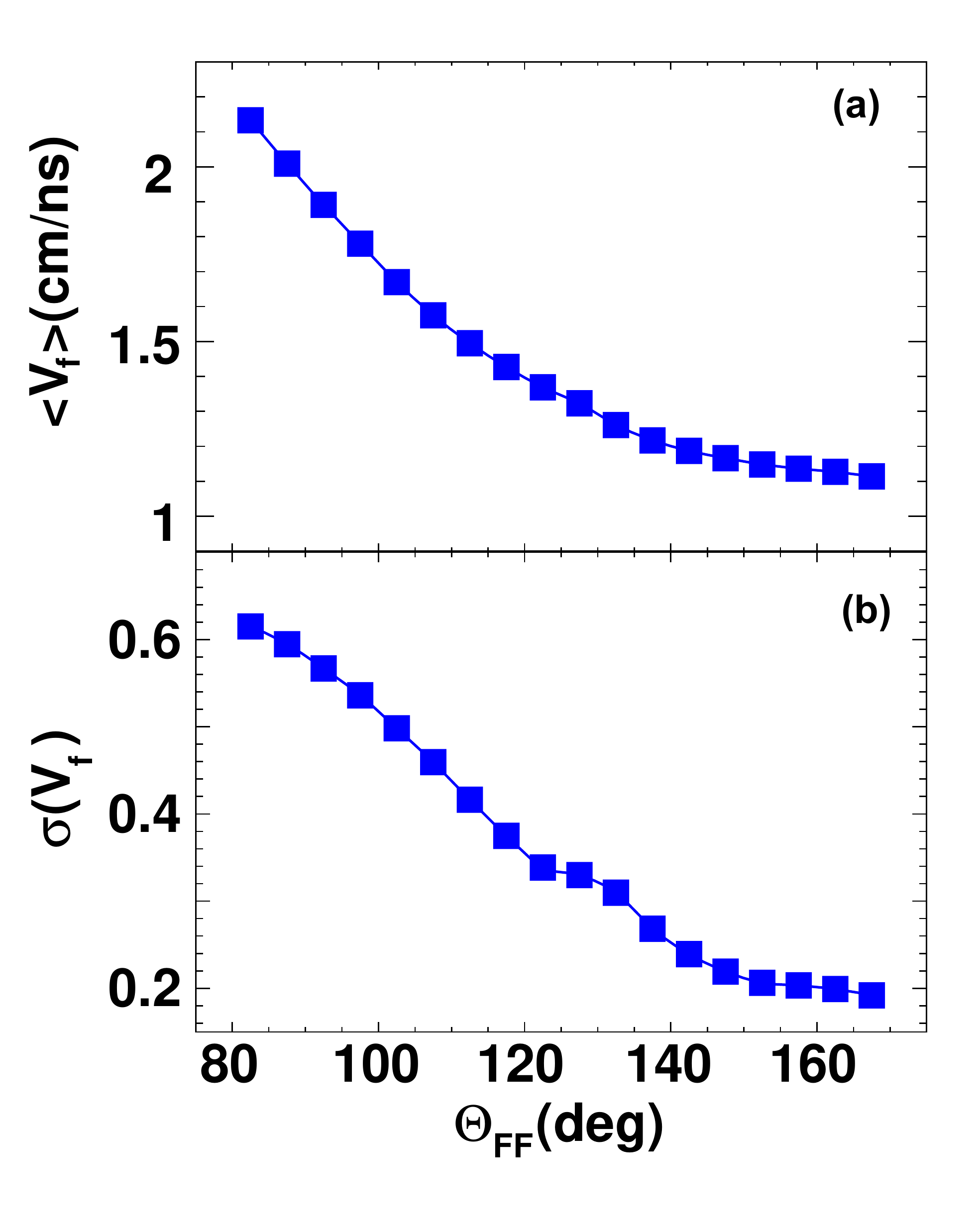}
\caption{(Color online) The average value of velocity   (a) and its standard deviation (b) of the velocity distribution  as a function of  folding angle.}
\label{v_folding}
\end{figure}

 The dynamic feature of the fast fission can be further explored from the velocity distribution of the FFs. We present in Fig. \ref{v_folding} the average (a) and the standard deviation (b) of the  velocities of the FFs recorded in the PPACs as a function of folding angle. From panel (a), it can be seen that the average velocity value $\left<v_{\rm f}\right>$  decreases with the folding angle.  Panel (b) presents the standard deviation of the velocity $\sigma(v_{f})$, and it is also clear that the broadening of the velocity of the FFs decreases with the folding angle.  This result is consistent with the picture seen in the trend of $\sigma(\Delta \phi)$ in figure \ref{d_phi}.   The scission point is early reached when the excitation energy of the fissioning TLF is still high, thus,  the statistical fluctuation (corresponding to the left excitation energy) enhances the variance of the velocity of the FF. This is consistent with the earlier experimental observation in ${\rm Ar+^{209}Bi}$  reactions at 25 MeV/u \cite{zjw1999,zjw1999-2}.

Finally, the dynamic feature of the fast fission may also cause the anisotropy of the angular distribution of the fission axis. Here the fission axis is defined as the vector of the relative velocity $\vec{v}_{FF}$ from the $\rm f_2$ to $\rm f_1$, where  $\rm f_1$ and  $\rm f_2$ are the two fragments. The angle of the fission axis $\Lambda_{\rm FF}$ is the angle of the fission axis w.r.t. the beam axis, as defined in \cite{Wuqh2019}. Usually, the experimental measurement of the distribution of   $\Lambda_{\rm FF}$ requires fine correction of the geometric efficiency, and hence it is better feasible using $4\pi$ detecting system. 

\begin{figure}[!htb]
\includegraphics[width=.35\textwidth]{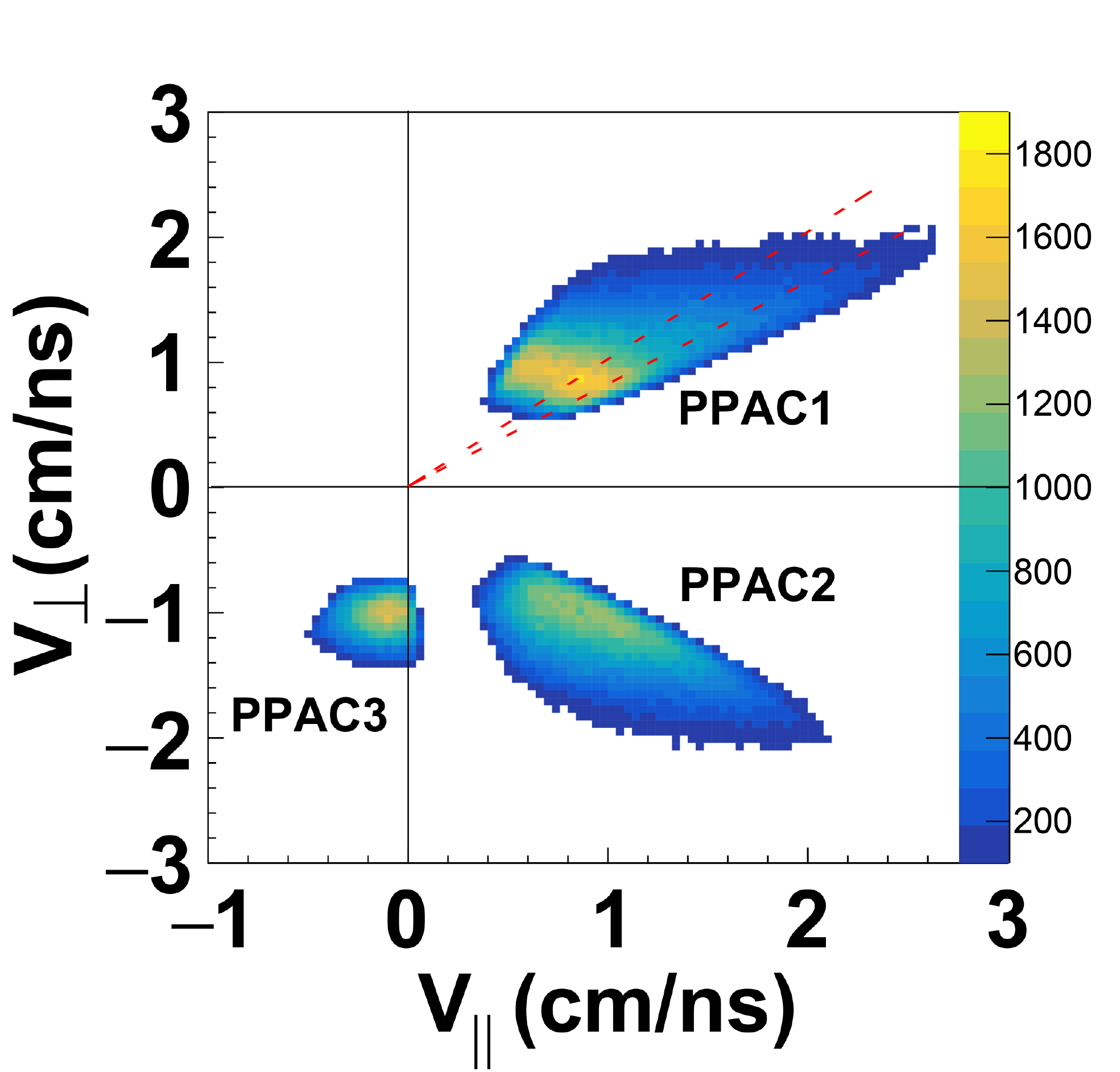}
\caption{(Color online) The transverse and parallel velocity distribution of the fission fragments.}
\label{v_pp}
\end{figure}

In our experiment, PPACs cover only part of the whole space, in order to introduce less ambiguity to the geometry efficiency correction, we fix the direction of the first FF in PPAC1. In this case, we need only to correct the efficiency of the second FF on PPAC 2 and PPAC3, and the trend of the angular of the fission axis can be inferred. Figure \ref{v_pp} presents  the correlation plot of  parallel and transverse velocity of the FFs. Here the transverse  velocity of the FFs recorded in PPAC1 is defined as positive and that in other two PPACs are defined as negative. It is clearly shown there is a dead area of less than $20^\circ$ between PPAC2 and PPAC3. The dashed lines define a narrow range of $40^\circ<\theta_{\rm f_1}<45^\circ$, which is fixed in the investigation of the distribution properties of the fission axis as following. 

\begin{figure}[!htb]
\includegraphics[width=.5\textwidth]{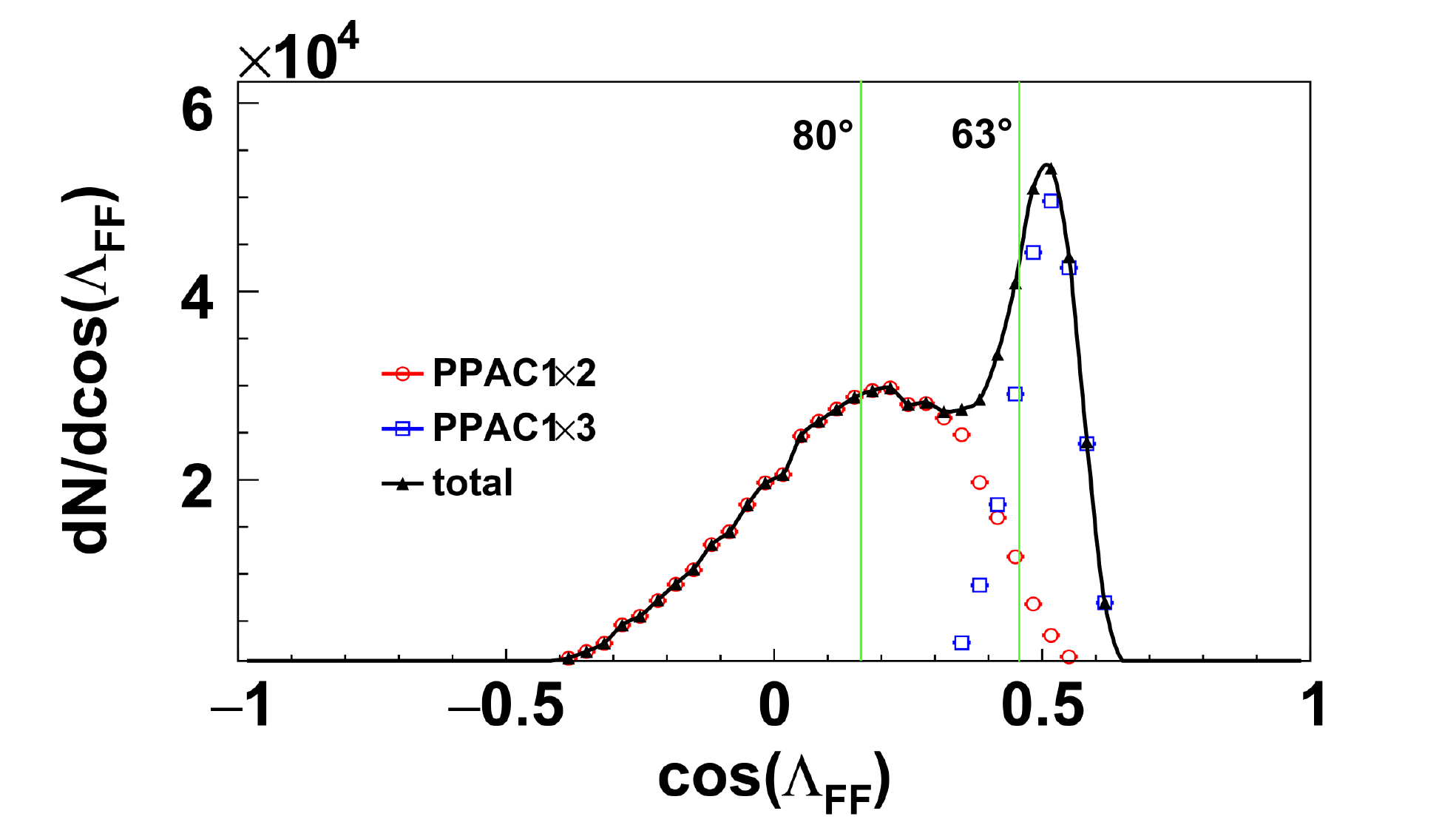}
\caption{(Color online) The angular distribution  $dN/d\cos (\Lambda_{\rm FF})$ of the fission axis w.r.t. the beam.}
\label{lambda}
\end{figure}

The  distribution of $d\sigma/d\cos (\Lambda_{\rm FF})$ is plotted in Fig. \ref{lambda}. The geometric efficiency arising from the incomplete azimuth coverage of PPAC2 and PPAC 3 is corrected in each $\Lambda_{\rm FF}$ bin. The events of \pacab~ and \pacac~ are represented by  symbols, while  the curve represents the sum. It is clearly shown that the deficiency of the gap between PPAC 2 and PPAC 3 causes a kink in a wide range in $63^\circ<\Lambda_{\rm FF}<80^\circ$. Regardless of the kink area and the uncovered region within $\Lambda_{\rm FF}<50^\circ$, it is shown that the distribution of $dN/d\cos (\Lambda_{\rm FF})$ increases steadily  with  $\cos(\Lambda_{\rm FF})$ and tends to peak at forward angle, which is at variance with the expectation of isotropic distribution for statistic fission. The trend is in qualitative agreement with the previously reported experimental results in HIR at Fermi energies \cite{Bocage2000,Filippo2005}.  

\section{Summary}\label{sec.V}

In summary, the fission fragments in  25 MeV/u \krpb~ reactions have been measured with CSHINE detection system.  At current phase  three PPACs and four SSDTs are mounted to measure the fission fragments and the coincident LCPs and IMFs, respectively. Using the timing and position signals of the PPACs and the start timing from the RF of the accelerator, one can measure the velocities of the fission fragments and reconstruct the fission events, where the linear momentum transfer can be derived  by the folding angle. It is shown that, the  width of the azimuthal angle difference, the mean value and the width of the velocity distribution of the fission fragments decrease  with the folding angle. The  anisotropic angular distribution of fission axis has been observed. The results are consistent with the picture that fast fission occurs.  Prospectively, with the ability of  reconstructing the fission events, CSHINE provides the opportunities in the studies of isospin dynamics and nuclear symmetry energy  by further counting  the coincident isotope-resolved LCPs and IMFs.  

\section{Acknowledgments}\label{sec.V}

This work has been supported by the National Natural Science Foundation of China under Grant Nos. 11875174 and 11961131010 and by the Polish National Science Center under Grant No. 2018/30/Q/ST2/00185.  We acknowledge the crystal group from IMP, CAS for providing the CsI detectors, the RIBLL group for offering local help in experiment and the machine staff for delivering the Krypton beam.


\begin{thebibliography} {9}



\bibitem{Liba2021} Bao-An Li et al., Universe, 7, 182 (2021). 
\bibitem{Ligo2017} B. P. Abbott et al. for LIGO collaboration, Phys. Rev. Lett. 119, 161101 (2017).
\bibitem{Ligo2018} B. P. Abbott et al. for LIGO collaboration, Phys. Rev. Lett. 121, 161101 (2018).

\bibitem{Tsang2004} M. B. Tsang et al, Phys. Rev. Lett.  92, 062701 (2004).
\bibitem{Chenlw2005} L. W. Chen et al., Phys. Rev. Lett. 94, 032701 (2005).
\bibitem{Tsang2001} M. B. Tsang  et al. Phys. Rev. Lett.  86, 5023 (2001).
\bibitem{Chenlw2010} L. W. Chen et al., Phys. Rev. C 82, 024321 (2010). 
\bibitem{Zz2014}  Z. Zhang and L. W. Chen, Phys. Rev. C  90, 064317 (2014).
\bibitem{Tsang2009} M. B. Tsang et al.,  Phys. Rev. Lett. 102, 122701  (2009).
\bibitem{Zy2017} Y. Zhang et al., Phys. Rev. C 95, 041602(R) (2017).

\bibitem{Prex2} D. Adhikari et al.,  Phys. Rev. Lett. 126, 172502 (2021).
\bibitem{Reed2021} B. T. Reed et al,   Phys. Rev. Lett. 126, 172503 (2021).
\bibitem{Sprit2021} J. Estee et al.,   Phys. Rev. Lett. 126, 162701 (2021).
\bibitem{Yongjia2020} Y. J. Wang and Q. F. Li, Frontiers of Physics, 15, 1 (2020).
\bibitem{LLM2017} L. M. Lyu et al., Sci. China-Phys. Mech. Astron. 60, 012021 (2017).
\bibitem{LLM2020}L. M. Lyu et al., Nucl. Sci. Tech. 31, 11 (2020).

\bibitem{Lorusso2015} G. Lorusso, et al., Phys. Rev. Lett. 114,  192501 (2015).
\bibitem{Nishimura2012} N. Nishimura, et al., Phys. Rev. C 85, 048801 (2012).
\bibitem{Suzuki2012} T. Suzuki, et al., Phys. Rev. C 85, 015802 (2012).


\bibitem{PBR96} K.Pomorski et al., Nuclear Physics A 605, 87 (1996).
\bibitem{SCH99} P. Schuurmans et al., Phys. Rev. Lett 82, 4787 (1999).
\bibitem{JB2007} Y. Jia and J.D. Bao,  Phys. Rev. C 75, 034601 (2007).
\bibitem{LB2011} Z. H. Liu and J. D. Bao, Phys. Rev. C 83, 044613 (2011).
\bibitem{Zhanghf2014} H. F. Zhang et al., Phys. Rev. C 90, 054313 (2014).
\bibitem{Tanimura2017} Y. Tanimura et al., Phys. Rev. Lett 118, 152501 (2017).
\bibitem{TZL17} H. Tao et al., Phys. Rev. C 96, 024319 (2017).
\bibitem{WY2018} N. Wang and W. Ye,  Phys. Rev. C 97, 014603 (2018).
\bibitem{WY2018-2} N. Wang and W. Ye, Phys. Rev. C 98, 034614 (2018).
\bibitem{Pomorski2021} K. Pomorski et al., Chin. Phys. C 45, 054109 (2021). 
\bibitem{Pavel2021} Pavel V. Kostryukov et al., Chin. Phys. C 45, 124108 (2021).
\bibitem{Zhengh2018} H. Zheng et al., Phys. Rev.  C 98, 024622 (2018).
\bibitem{Guol2018} L. Guo et al.,   Phys. Rev. C  98, 064609 (2018).


\bibitem{Greg82} C. Gregoire  et al., Nucl. Phys. A 387, 37 (1982).
\bibitem{Greg82t} C. Gregoire  et al., Nucl. Phys. A 383, 392 (1982).
\bibitem{Gla83} P. Gl\"aissel et al., Z. Phys. A  310, 189 (1983).
\bibitem{Leray84} S. Leray et al., Nucl. Phys. A 423, 175 (1984).
\bibitem{Zheng84} Z. Zheng et al., Nucl. Phys A 422, 447 (1984).

\bibitem{Wen13} Kai Wen et al., Phys. Rev. Lett. 111, 012501 (2013).
\bibitem{Russ11} P. Russotto et al.,  Phys. Lett. B 697, 471 (2011).
\bibitem{Riz11} C. Rizzo et al, Phys. Rev. C 83, 014604 (2011).
\bibitem{TL09} Junlong Tian, X. Li et al., Eur. Phys. J. A 42, 105 (2009).
\bibitem{TO11} Junlong Tian, Li Ou et al., Int. J. Mod. Phys. E, 20, 1755 (2011).
\bibitem{LTQ13} Li Cheng, Junlong Tian et al., Chin. Phys. C 37, 114101 (2013).
\bibitem{TW08} Junlong Tian, Xizhen Wu et al., Phys. Rev. C 77, 064603 (2008).
\bibitem{LTO13} Cheng Li, Junlong Tian, Li Ou et al., Phys. Rev. C 87, 064615 (2013).
\bibitem{WT11} Ning Wang, Junlong Tian et al., Phys. Rev. C 84, 061601(R) (2011).
\bibitem{God2015} P. Goddard et al., Phys. Rev. C 92, 054610 (2015).

\bibitem{Bocage2000} F. Bocage et al., Nucl. Phys. A 676, 391 (2000).
\bibitem{Filippo2005} E. De Filippo et al., Phys. Rev. C 71, 064604 (2005).
\bibitem{Filippo2012} E. De Filippo et al., Phys. Rev. C 86, 014610 (2012).
\bibitem{Pagano2018} E.V. Pagano et al., Jour. of Phys. Conf. Series 1014, 012011 (2018). 
\bibitem{Piantelli2020} S. Piantelli et al., Phys. Rev. C 101, 034613 (2020).
\bibitem{WRS2014} R. S. Wang et al., Phys. Rev. C 89, 064613 (2014).
\bibitem{Casini1993} G. Casini et al., Phys. Rev. Lett. 71, 2567 (1993).

\bibitem{Wuqh2019} Q. Wu et al., Phys. Lett. B 797, 134808 (2019).
\bibitem{Wuqh2020} Q. Wu et al., Phys. Lett. B 811, 135865 (2020).
\bibitem{Pei2020} M. Pancic et al.,  Front. Phys. 8, 351 (2020).

\bibitem{Guanfh2021} F. H. Guan et al., Nucl. Inst. Meth.  A, 1011, 165592 (2021).
\bibitem{Wangyj2021} Y. J. Wang et al., Nucl. Sci. Tech. 32, 4 (2021). 
\bibitem{Guanfh2022} F. H. Guan et al.,  arXiv:2110.08261v1.
\bibitem{Wangyj2022} Y. J. Wang et al.,  10.1016/j.physletb.2021.136856.

\bibitem{Weixl2020} X. L. Wei et al., Nucl. Eng. Tech. 52, 575 (2020).

\bibitem{Viola} V. E. Viola, Phys. Rev. C 31 1550 (1985).

\bibitem{zjw1999}   J. W. Zheng et al., High Ene. Phys. Nucl. Phys. 23, 409 (1999).
\bibitem{zjw1999-2} J. W. Zheng et al., High Ene. Phys. Nucl. Phys. 23, 946 (1999).

    
\end{thebibliography}
\end{document}